\documentclass[11pt]{article}
\usepackage{epsfig}
\usepackage{amssymb}
\usepackage{tweaklist}
\usepackage{enumerate}

\renewcommand{\enumhook}{\setlength{\topsep}{0pt}%
  \setlength{\itemsep}{0pt}}

\newcommand{\lastcorrections}%
{{\vskip 0.2in
\begin{sloppypar}
    \baselineskip -0.2in
\ \\
    \tiny\bf\noindent Last Corrections:\\   
work33.tex\\
wolf: Sun Nov  2 15:44:26 PST 2003.  This is the last version of work05.\\
larry: Fri Jan  2 20:05 MET 2004.\\
larry: Sun Jan  4 16:05:10 MET 2004.  This is the last version of work06.\\
larry: Mon Jan  5 08:36:09 MET 2004.  This is the last version of work07.\\
larry: Mon Jan  5 15:12:12 MET 2004.  This is the last version of work08.\\
wolf:  Mon Jan  5 23:55:17 MET 2004.  This is the last version of work09.\\
larry: Tue Jan  6 03:49:32 MET 2004.\\
wolf:  Wed Jan  7 09:07 MET 2004.  This is the last version of work10.\\
larry: Thu Jan  8 13:57:30 MET 2004. This is the last version of work11.\\
wolf:  Thu Jan  8 21:33 MET 2004.  This is the last version of work12.\\
larry: Fri Jan  9 07:22:04 MET 2004.\\
wolf:  Fri Jan  9 09:44:02 MET 2004.\\
wolf: Sat Mar 20 22:21:39 PST 2004.\\
wolf: Mon Apr  5 02:05:41 PDT 2004.  This is the last version of work15.\\
work16.tex\\
larry: Tue Apr  6 14:04:23 PDT 2004.  This is the last version of work16.\\
work17.tex\\
larry: Tue Apr  6 14:04:23 PDT 2004 This is the last version of work17.\\
work18.tex\\
wolf: Tue Apr  6 23:05:55 PDT 2004.  This is the last version of work18.
work19.tex\\
larry: Wed Apr  7 03:11:59 PDT 2004.\\
larry: Wed Apr  7 06:33:37 PDT 2004.  This is the last version of work19.\\
work20.tex\\
wolf: Wed Apr  7 12:52:05 PDT 2004.\\
larry: Wed Apr  7 13:35:23 PDT 2004.\\
wolf: Wed Apr  7 17:05:20 PDT 2004\\
wolf: Thu Apr  8 11:56:16 PDT 2004 work23, Rejected for FOCS 2004\\
wolf: Wed Jun  30 1:46:16 PDT 2004 work25, Submitted to SODA 2004\\
wolf: Wed Jun  30 1:46:16 PDT 2004 work26, Submitted to SODA 2004\\
work28.tex\\
larry: Fri Sep 10 09:26:47 PDT 2004\\
work29.tex\\
larry: Tue Sep 14 02:40:40 PDT 2004\\
larry: Tue Sep 14 06:06:16 PDT 2004\\
larry: Wed Sep 15 09:46:41 PDT 2004\\
wolf: Thu Sep 16 13:16:05 PDT 2004 work30.tex, last version with all tables\\
wolf: Fri Sep 17 23:14:50 PDT 2004 work31.tex\\
wolf: Mon Sep 27 13:57:45 PDT 2004 work33.tex\\
wolf: Mon Jan 22 02:34:42 PST 2007

\end{sloppypar}
}}

\newcommand{\wolfcomment}[1]%
    {{%
      \marginpar{{\tiny\begin{minipage}{0.5in}
                       \begin{flushleft}
                          {#1}
                       \end{flushleft}
                       \end{minipage}
                }}
    }}

\newcommand{\margincomment}[1]%
{{%
      \marginpar{{\tiny\begin{minipage}{0.5in}
                       \begin{flushleft}
                          {#1}
                       \end{flushleft}
                       \end{minipage}
                }}
    }}

\newcommand{\offset}{{\mbox{\it adjust}}}
\newcommand{\etal}{{\em et al.~\/}}
\newcommand{\tH}{{\rm th}}

\newcommand{\etc}{{\em etc.\/}}
\newcommand{\ie}{{\em i.e.\/}}

\newcommand{\myparagraph}[1]{\medskip{\noindent{\bf #1}}}

\newcommand{\calA}{{\cal A}}

\newcommand{\calM}{{\cal M}}

\newcommand{\calP}{{\cal P}}

\newcommand{\calR}{{\cal R}}

\newcommand{\calX}{{\cal X}}

\newcommand{\maly}[1]{{\scriptscriptstyle#1}}
\newcommand{\malyz}{{\maly{0}}}
\newcommand{\malyzero}{{\maly{0}}}
\newcommand{\malyone}{{\maly{1}}}

\newcommand{\parend}[1]{{\left( #1  \right) }}

\newcommand{\braced}[1]{{\left\{ #1  \right\} }}

\newcommand{\leftbracedthree}[3]{\left\{\begin{array}{l}
                                        #1 \\ #2 \\ #3
                                    \end{array}\right.
                           }

\newcommand{\mycase}[1]{\smallskip\noindent \mbox{Case #1:}}

\newcommand{\ignore}[1]{}

\newtheorem{theorem}{Theorem}

\newtheorem{corollary}{Corollary}

\newtheorem{lemma}{Lemma}

\newenvironment{proof}{{\noindent \it Proof:\/}}{$\square$}

\newcommand{\half}{{\mbox{$\frac{1}{2}$}}}
\newcommand{\onehalf}{{\half}}
\newcommand{\onethird}{{\mbox{$\frac{1}{3}$}}}
\newcommand{\twothirds}{{\mbox{$\frac{2}{3}$}}}

\newcommand{\onefourth}{{\mbox{$\frac{1}{4}$}}}
\newcommand{\onefifth}{{\mbox{$\frac{1}{5}$}}}

\newcommand{\threefifths}{{\mbox{$\frac{3}{5}$}}}

\newcommand{\threefourths}{{\mbox{$\frac{3}{4}$}}}
\newcommand{\fivefourths}{{\mbox{$\frac{5}{4}$}}}

\newcommand{\onesixth}{{\mbox{$\frac{1}{6}$}}}
\newcommand{\oneeighth}{{\mbox{$\frac{1}{8}$}}}
\newcommand{\fivesixths}{{\mbox{$\frac{5}{6}$}}}
\newcommand{\fivethirds}{{\mbox{$\frac{5}{3}$}}}

\newcommand{\elevensixths}{{\mbox{$\frac{11}{6}$}}}

\newcommand{\threehalves}{{\mbox{$\frac{3}{2}$}}}

\newcommand{\onetenth}{{\mbox{$\frac{1}{10}$}}}

\newcommand{\update}{{\wedge}}
\newcommand{\opt}{{\mbox{{\small \it opt}}}}

\newcommand{\cost}{{\mbox{\it cost}}}

\newcommand{\malyopt}{{\mbox{\tiny\it opt}}}

\newcommand{\reals}{{\bf R}}

        {\begin{list}{$-$}{\setlength{\itemsep}{-0.05in}}%
        }%
        {\end{list}}

\title{Knowledge State Algorithms:  Randomization with Limited Information}

\author{Wolfgang W. Bein
        \thanks{Department of Computer Science,
 Center for the Advanced Study of Algorithms,
        University of Nevada,
        Las Vegas, NV 89154.
         Email: {\tt bein@cs.unlv.edu}.
 Research supported by NSF grant CCR-0312093.
           }
        \and
        Lawrence L. Larmore
        \thanks{Department of Computer Science,
 Center for the Advanced Study of Algorithms,
        University of Nevada,
        Las Vegas, NV 89154.
        Email: {\tt larmore@cs.unlv.edu}.
 Research supported by NSF grant CCR-0312093.
               }
	\and
	R\"udiger Reischuk
	\thanks{Institut f\"ur Theoretische Informatik,
	Universit\"at L\"ubeck,
	Wallstra\ss{}e 40,
	D-23560 L\"ubeck}.
        }

\begin{document}

\maketitle

\thispagestyle{empty}

%
%
%



\begin{abstract}
We introduce the concept of knowledge states; many well-known 
algorithms can be viewed as knowledge state algorithms. 
The knowledge state approach can be used to
to construct competitive randomized online algorithms and 
study the tradeoff between competitiveness and memory.
A  knowledge state simply states conditional obligations
of an adversary, by fixing a work function, and gives a distribution
for the algorithm.  When a knowledge state algorithm receives a request,
it then calculates one or more ``subsequent" knowledge states,
together with a  probability of transition to each.  The algorithm
then uses randomization to select one of those subsequents to be the
new knowledge state.  
We apply the method to the paging problem. We present optimally
competitive algorithm for paging for the cases where the cache sizes 
are $k=2$ and $k=3$. These algorithms use only a very limited 
number of bookmarks.


\end{abstract}

\section{Motivation and Background}\label{sec: background}

In this paper we introduce a new method for constructing randomized
online algorithms, which we call the {\em knowledge state\/} model.
The purpose of this method is the address the trade-off between
memory and competitiveness.
The model is introduced and fully described for the first time in this
publication, but we note that a number of published algorithms
are implicitly consistent with the model although not in its full power.
For example, the algorithm EQUITABLE \cite{AcChNo00} is a
knowledge state algorithm for the $k$-cache problem that achieves the
optimal randomized competitiveness of $H_k$ for each $k$, using only
$O(k^2\log k)$ memory, as opposed to the prior algorithm, PARTITION
\cite{KouPap94B}, that uses the full information contained in the
work function, and hence requires unlimited memory as the length of the
request sequence grows.
At the other end of the scale, the randomized algorithm RANDOM\_SLACK
\cite{CoDoRS90} is in fact an extremely simple knowledge state
algorithm, which achieves randomized 2-competitiveness
for the 2-server problem for all metric spaces, and which
achieves randomized $k$-competitiveness for the
$k$-server problem on some spaces, including trees. We also
note that RANDOM\_SLACK  is {\em trackless\/} and is
an order 1 knowledge state algorithm, \ie, its  distribution
is supported by only one state.
(See the recent ACM SIGACT column \cite{sigact04} for a summary
of tracklessness; see also \cite{BeFlLa00,BeiLar00,BaChLa00,BeLaRe04}.)
We also note that we have recently
used the knowledege state technique to develop an optimally
competitive algorithm for the caching problem in
shared memory multiprocessor systems \cite{BeLaRe04}.

It is still an open question, whether there exists an
optimally competitive order $O(k)$ bookmark
randomized algorithm for the $k$-cache problem. An affirmative
answer to this question would settle an open problem listed in
\cite{BorElY98}. In this paper we describe progress on this question.
We give an order 2 knowledge state algorithm which is
provably $H_2$-competitive.  Since an equivalent behavioral
algorithm must keep one ``bookmark," namely the address of an ejected page,
it is not an improvement over our earlier result \cite{BeFlLa00},
but it does illustrate
the knowledge state technique in a simple way.
We then give an order 3 knowledge state algorithm which is
provably $H_3$-competitive,  which is an improvement, in terms of
memory requirements, over EQUITABLE
for the case $k=3$ (Section \ref{sec: ks cache}).

We also consider
the problem of breaking the 2-competitive barrier for
the randomized competitiveness of the 2-server problem, a goal which has,
as yet, been achieved only in special cases (Section \ref{sec: server}).
For the class of uniform
spaces, this barrier was broken by PARTITION \cite{KouPap94B}.
For the line, a $\frac{155}{78}$-competitive algorithm was given
by Bartal {\etal}\cite{BaChLa00}.  

In this paper we give a formal description of the knowledge state
method. It is defined  using the mixed  
model of online computation, which is described in Section 
\ref{sec: online opt prob}.  This section relates the mixed model
to the standard models of online computation, and explains how 
a behavioral algorithm can be derived from a mixed model description.
Section \ref{sec: know} defines the knowledge state method (in terms
of the mixed model) and shows how potentials can be used to 
derive the competitive ratio of a knowledge state algorithm.  
Even though the concepts in  Section
\ref{sec: online opt prob} and \ref{sec: know}  are natural and 
intuitive some of the formal arguments to prove our method
are somewhat involved.
In Section \ref{sec: ks cache} the method is applied to the 
paging problem; two optimally competitive 
algorithms are presented.
We  discuss ongoing experimental work  for the server problem
in Section \ref{sec: server}.

\section{The Mixed Model of Online Computation}\label{sec: online opt prob}
We will introduce a new model of randomized online computation which
is a generalization of both the classic behavioral and distributional
models.
We assume that we are given an online problem with states  $\calX$
(also called configurations),
a fixed {\em start state\/} $x^\malyz\in\calX$, and a requests $\calR$.
If the current state is
$x\in\calX$ and a request $r\in\calR$ is given, an algorithm for the
problem must {\em service\/} the request by choosing a new state $y$
and paying a cost, which we denote $\cost(x,r,y)$.
It is convenient to assume that there is a ``distance" function $d$ on
$\calX$, and it is possible to choose to move from state $x$ to state $y$
at cost $d(x,y)$ at any time, given no request.  We will assume that
$d(x,x)=0$ and $d(x,z)\le d(x,y)+d(y,z)$ for any states $x,y,z$.
It follows that $\cost(u,r,v)\le d(u,x)+cost(x,r,y)+d(y,v)$
for any states $u,x,y,v$ and request $r$. Formally in this paper we refer to
an online problem as an ordered triple $\calP=(\calX,\calR,d)$.
Examples of online problems satisfying these conditions abound, such
as the server problem, the cache problem, \etc.

Given a {\em request sequence} $\varrho=r^\malyone, \ldots r^n$, an algorithm
must choose a sequence of states $x^\malyone, \ldots x^n$,
the {\em service\/}.  The {\em cost\/} of this service is defined
to be $\sum^n_{t=1}\cost(x^{t-1},r^t,x^t)$.
An {\em offline\/} algorithm knows $\varrho$ before choosing the
service sequence, while an {\em online\/} algorithm must choose $x^t$
without knowledge of the future requests.
We will assume that there is an optimal offline algorithm, $\opt$, which
computes an optimal service sequence for any given request sequence.
As is customary we say that a deterministic
online algorithm $\calA$ is {\em $C$-competitive\/}
for a given number $C$
if there exists a constant $K$ (not dependent on $\varrho$)
such that
$\cost_\calA(\varrho)\le C\cdot\cost_\malyopt(\varrho)+K$
for any request sequence $\varrho$.
Similarly, we say that a randomized
online algorithm $\calA$ is $C$-competitive for a given number $C$
if there exists a constant $K$ (not dependent on $\varrho$) such that
$E\parend{\cost_\calA(\varrho)}\le C\cdot\cost_\malyopt(\varrho)+K$
for any request sequence $\varrho$, where $E$ denotes expected value.

In order to make the description of various models  of randomized online
computation more precise, we introduce the following notation.
Let $\Pi$ be the set of all finite distributions on $\calX$.
If $\pi\in\Pi$ and $S\subseteq\calX$, we say that $S$ {\em supports \/}
the distribution $\pi$ if $\pi(S)=1$.  The {\em distributional support\/}
(or ``{\em support\/}" for short) of any $\pi\in\Pi$ is defined to be the
unique minimal set which supports $\pi$.  By an abuse of notation, if 
the support of $\pi$ is a singleton $\braced{x}$, we write $\pi=x$.

An instance of the {\em transportation problem\/} is a weighted directed
bipartite graph with distributions on both parts.
Formally, an instance is an ordered quintuple
 $(A,B,\cost,\alpha,\beta)$ where $A$ and $B$ are finite non-empty sets,
$\alpha$ is a distribution on $A$, $\beta$ is a distribution on $B$,
and $\cost$ is a real-valued function on $A\times B$.
A {\em solution\/} to this instance is a distribution $\gamma$ on $A\times B$
such that
\begin{enumerate}
\item
$\gamma(\braced{a}\times B) = \alpha(a)$ for all $a\in A$.
\item
$\gamma(A\times\braced{b}) = \beta(b)$ for all $b\in B$.
\end{enumerate}
Then $\cost(\gamma)=\sum_{a\in A}\sum_{b\in B}\gamma(a,b)\cost(a,b)$,
and $\gamma$ is a {\em minimal\/} solution if $\cost(\gamma)$ is minimized
over all solutions, in which case we call $\cost(\gamma)$ the
{\em minimum transportation cost}.

There are three standard models of randomized online algorithms (see, for
example \cite{BorElY98}).  We introduce a new model in this paper, which we
call the {\em mixed model}.  Those three standard models are:
distribution of deterministic online algorithms, the behavioral model,
and the distributional model.
We very briefly describe the three standard models.

\myparagraph{Distribution of Deterministic Online Algorithms.}
In this model, $\calA$ is a random variable whose value is a deterministic
online algorithm.  If the random variable has a finite distribution, we say
that $\calA$ is {\em barely random\/}.

\myparagraph{Behavioral Online Algorithms.}
In this model $\calA$ uses randomization at each step to pick the next
configuration.  We assume that $\calA$ has memory.  Let $\calM$ be the
set of all possible memory states of $\calA$.
We define a {\em full state\/} of $\calA$ to be an ordered pair
 $k=(x,m)\in\calX\times\calM$.
Let $m^\malyz\in\calM$ be the initial memory state, and let $m^t$ be the
memory state of $\calA$ after servicing the first $t$ requests.

Then $\calA$ uses randomization to compute $k^t=(x^t,m^t)$,
 the full state after $t$ steps,
given only $k^{t-1}$ and $r^t$.
A behavioral algorithm can then be thought of as a function on
 $\calX\times\calM\times\calR$
 whose values are random variables in $\calX\times\calM$.

\myparagraph{Distributional Online Algorithms.}
If $\pi,\pi'\in\Pi$,
let $S$ be the support of $\pi$ and
$S'$ be the support of $\pi'$.
We then define
$d(\pi,\pi')$ to be the minimum transportation cost of the transportation
problem $(S,S',d,\pi,\pi')$, and if
  $r\in\calR$, we define $\cost(\pi,r,\pi')$ to be the
minimum transportation cost of the transportation
problem $(S,S',cost^r,\pi,\pi')$,
 where $\cost^r=\cost(\ ,r,\ ):\calX\times\calX\to\reals$.

A distributional online algorithm
 $\calA$ is then defined as follows.
\begin{enumerate}
\item
There is a set $\calM$ of memory states of $\calA$.
There is a start memory state $m^\malyz\in\calM$.

\item
A {\em full state\/} of $\calA$ is a pair $k=(\pi,m)\in\Pi\times\calM$.
The initial full state is $k^\malyz=(\pi^\malyz,m^\malyz)$,
where $\pi^\malyz=s^\malyz$.
\item
For any given full state $k=(\pi,m)$ and request $r$, $\calA$
deterministically computes a new full state $k'=(\pi',m')$, using only
the inputs $\pi$, $m$, and $r$.  We write $\calA(\pi,m,r)=(\pi',m')$ or
alternatively $\calA(k,r)=k'$.  Thus, $\calA$ is a function from
$\Pi\times\calM\times\calR$ to $\Pi\times\calM$.

\item
Given any input sequence $\varrho=r^1\ldots r^n$, $\calA$ computes
a sequence of full states $\calA(\varrho) = k^\malyone,\ldots k^n$,
following the rule that $k^t=(\pi^t,m^t) = \calA(k^{t-1},r^t)$ for
all $t \ge 1$.  Define
$\cost_\calA(\varrho) = \sum^n_{t=1}\cost(\pi^{t-1},r^t,\pi^t)$.
\end{enumerate}

We note that a distributional online algorithm, despite being a model for
a randomized online algorithm, is in fact deterministic, in the sense that
the full states $\braced{k^t}$ are computed deterministically.

The following theorem is well-known. (It is, for example, implicit in
Chapter 6 of \cite{BorElY98}.)
\begin{theorem}\label{thm: all models equivalent}
All three of the above models of randomized online algorithms are equivalent,
in the following sense.  If $\calA_1$ is an algorithm of one of the models,
there exist algorithms $\calA_2$, $\calA_3$, of each of
the other models, such that, given any request sequence $\varrho$,
the cost (or expected cost) of each $\calA_i$ for $\varrho$ is no
greater than the cost (or expected cost) of $\calA_1$.
\end{theorem}


\myparagraph{The Mixed Model.}
The {\em mixed model\/} of randomized algorithms is a generalization of
both the behavioral model and the distributional model.  A mixed
online algorithm chooses a distribution at each step, but, as opposed to
a distributional algorithm, which must make that choice deterministically,
can use randomization to choose the distribution.

A {\em mixed\/} online algorithm $\calA$ for an online problem
$\calP = (\calX,\calR,d)$ is defined as follows.  As before,
let $\Pi$ be the set of finite distributions on $\calX$.
\begin{enumerate}
\item
There is a set $\calM$ of memory states of $\calA$.
There is a start memory state $m^\malyz\in\calM$.
\item
A {\em full state\/} of $\calA$ is a pair $k=(\pi,m)\in\Pi\times\calM$.
The initial full state is $k^\malyz=(\pi^\malyz,m^\malyz)$,
where $\pi^\malyz=s^\malyz$.
\item
For any given full state $k=(\pi,m)$ and request $r$,
there exists a finite set of full states $k_\malyone,\ldots k_m$
and probabilities $\lambda_\malyone\ldots \lambda_m$,
where $\sum^m_{i=1}\lambda_i=1$,
such that if the current full state is $k$ and the next request is $r$,
$\calA$ uses randomization to compute a new full state $k'=(\pi',m')$,
by selecting $k'=k_i$ for some $i$.
The probability that $\calA$ selects each given $k_i$ is $\lambda_i$.
We call the $\braced{k_i}$ the {\em subsequents\/} and
the $\braced{\lambda_i}$ the {\em weights\/} of the subsequents,
for the request $r$ from the full state $k$.

$\calA$ is a function on $\Pi\times\calM\times\calR$ whose values are
random variables in $\Pi\times\calM$.
We can write $\calA(\pi,m,r)=(\pi',m')$.
Alternatively, we write $\calA(k,r)=k'$.   For fixed $k$ and $r$;
$k', \pi'$, and $m'$ can be regarded as random variables.
\item
Given any input sequence $\varrho=r^1\ldots r^n$, $\calA$ computes
a sequence of full states $\calA(\varrho) = (\pi^1,m^1)\ldots (\pi^n,m^n)$,
following the rule that $k^t=(\pi^t,m^t) = \calA(k^{t-1},r^t)$
for all $t > 1$.
Note that, for all $t>0$, $k^t$, $\pi^t$, and $m^t$ are random variables.
\end{enumerate}

Computing the cost of a step of a mixed model online algorithm $\calA$
is somewhat tricky. We note that it might seem
that $\sum^m_{i=1}\lambda_i\cost(\pi,r,\pi_i)$
would be that cost; however, this is an overestimate.

Without loss of generality, $\calA$ is sensible.
Let $k=(\pi,m)\in\Pi\times\calM$ and let $r\in\calR$.
Let $S\subseteq \calX$ be the support of $\pi$.
Let $\braced{k_i=(\pi_i,m_i)}$ be the subsequents and $\braced{\lambda_i}$
the weights of the subsequents, for the request $r$ from the full state $k$.
Let $\bar S\subseteq\calX$ be the union of the supports of the
$\braced{\pi_i}$.
Define $\bar\pi=\sum^m_{i=1}\lambda_i\pi_i$.
Note that $\bar\pi\in\Pi$, and its support is $\bar S$.
Define $\cost_\calA(k,r) = \cost(\pi,r,\bar\pi)$.

Finally, if $\varrho=r^\malyone\ldots r^n$
is the input request sequence, and the sequence of full states
of $\calA$ is $k^\malyone\ldots k^n$,
we define $\cost_\calA(\varrho)=\sum^n_{t=1}\cost_\calA(k^{t-1},r^t)$.

We now prove that the mixed model for randomized online algorithms is
equivalent to the three standard models.

\begin{lemma}\label{lem: mixed yields beh}
If $\calA$ is a mixed online algorithm,
there is a behavioral online algorithm $\calA'$ such that,
for any request sequence $\varrho$,
$E\parend{\cost_{\calA'}(\varrho)} = E\parend{\cost_\calA(\varrho)}$.
\end{lemma}

\begin{proof}
A memory state of $\tilde\calA$ will be a full state of $\calA$,
\ie, we could write $\tilde\calM\subseteq\Pi\times\calM$.  By a slight
abuse of notation, we also define a full state of $\tilde\calA$ to be
an ordered triple $(x,\pi,m)\in\calX\times\Pi\times\calM$ such that
$(\pi,m)$ is a full state of $\calA$ and $\pi(x) > 0$.
Intuitively, $\tilde\calA$ keeps track of its true state $x\in\calX$, while
remembering the full state $(\pi,m)$ of an emulation of $\calA$.

For clarity of the proof, we introduce more complex notation for some of
the quantities defined earlier.
Let $\pi,\sigma\in \Pi$, $m,n\in \calM$, and $r\in\calR$.
If $(\pi,m)$ is a full state of $\calA$,
define $\lambda_{\pi,m,r,\sigma,n}$ to be the
probability that $\calA(\pi,m,r) = (\sigma,n)$, \ie, the
conditional probability that $\calA$ chooses $(\sigma,n)$
to be the next full state,
given that the current full state is $(\pi,m)$ and the request is $r$.
We assume that there can be at most finitely many choices of $(\sigma,n)$ for
which $\lambda_{\pi,m,r,\sigma,n} > 0$.
In case $(\pi,m)$ is not a full state of $\calA$,
then $\lambda_{\pi,m,r,\sigma,n}$ is
defined to be
zero.
If $(\pi,m)$ is a full state of $\calA$ and $r\in\calR$,
write $\bar\pi_{\pi,m,r} = \sum_{\sigma\in\Pi,n\in\calM}
\lambda_{\pi,m,r,\sigma,n}\cdot\sigma \in \Pi$, and
choose a finite distribution
$\gamma_{\pi,m,r}$ on $\calX\times\calX$
which is a minimal solution to the
transportation problem $\parend{\calX,\calX,\cost^r,\pi,\bar\pi_{\pi,m,r}}$,
where $\cost^r(x,y) = \cost(x,r,y)$.
Thus
$\pi(x) = \sum_{y\in\calX}\gamma_{\pi,m,r}(x,y)$
for $x\in\calX$;
$\bar\pi_{\pi,m,r}(y) = \sum_{x\in\calX}\gamma_{\pi,m,r}(x,y)$
for $y\in\calX$;
$\cost_\calA(\pi,m,r) = \sum_{x\in\calX,y\in\calX}\gamma_{\pi,m,r}(x,y)
\cost(x,r,y)$.

We now formally describe the action of the behavioral algorithm
$\tilde\calA$.
The initial full state of $\tilde\calA$ is $(x^\malyzero,k^\malyzero) =
(x^\malyzero,\pi^\malyzero,m^\malyzero)$.
Given that the full state of $\tilde\calA$ is $(x,\pi,m)$ and the next
request is $r\in\calR$, and given any
 $(y,\sigma,n)\in\calX\times\Pi\times\calM$, we define
$\Lambda_{x,\pi,m,r,y,\sigma,n}$, the probability that $\tilde\calA$
chooses the next full state to be $(y,\sigma,n)$, as follows:

\begin{description}
\item If $\bar\pi_{\pi,m,r}(y) = 0$, then
$\Lambda_{x,\pi,m,r,y,\sigma,n} = 0$.
\item Otherwise,
 $
\Lambda_{x,\pi,m,r,y,\sigma,n}
 =
\frac
{
\gamma_{\pi,m,r}(x,y)
\cdot
\sigma(y)
\cdot
\lambda_{\pi,m,r,\sigma,n}
}
{
\pi(x)
\cdot
\bar\pi_{\pi,m,r}(y)
}.
 $
\end{description}

Let $\varrho$ be a given request sequence.
We now prove that $E\parend{\cost_{\tilde\calA}(\varrho)}
= E\parend{\cost_\calA(\varrho)}$.
For any $t\ge 0$ and any knowledge state $(\pi,m)$ of $\calA$,
define $p^t(\pi,m)$ to be the probability that
the full state of $\calA$ is $(\pi,m)$ after $t$ steps.
Additionally, if $x\in\calX$, define $q^t(x,\pi,m)$ to be the probability
that the full state of $\tilde\calA$ is $(x,\pi,m)$ after $t$ steps.

\noindent
To prove the lemma we consider first the following two claims:
\begin{enumerate}
\item
For any $t\ge 0$, $x\in\calX$, $\pi\in\Pi$, and $m\in\calM$,
$q^t(x,\pi,m) = p^t(\pi,m)\cdot\pi(x)$.

\item
For any $t\ge 0$, $\pi\in\Pi$, and $m\in\calM$,
$\sum_{x\in\calX}q^t(x,\pi,m) = p^t(\pi,m)$.
\end{enumerate}

\noindent
We prove claims
1 and 2
by simultaneous induction on $t$.
If $t=0$, both claims are trivial by definition.
Now, suppose $t > 0$.
We verify claim 1 for $t$.
By the inductive hypothesis, claim 2 holds for $t-1$.
  Write $r=r^t$.
  Let $y,\sigma,n\in\calX\times\Pi\times\calM$.
If $(\sigma,n)$ is not a full state of $\calA$ or $\sigma(y)=0$,
we are done.  Otherwise, recall that
$\bar\pi_{\pi,m,r}(y) = \sum_{x\in\calX}\gamma_{\pi,m,r}(x,y)$
for all $y\in\calX$, and we obtain
{\small
\begin{eqnarray*}
q^t(y,\sigma,n)
&=&
\sum_{(x,\pi,m)\in\calX\times\Pi\times\calM}
q^{t-1}(x,\pi,m)\Lambda_{x,\pi,m,r,y,\sigma,n}\\
 &=&
\sum_{(x,\pi,m)\in\calX\times\Pi\times\calM,\pi(x)>0,\bar\pi_{\pi,m,r}(y)>0}
 p^{t-1}(\pi,m)\pi(x)\cdot
\frac{\gamma_{\pi,m,r}(x,y)\cdot\sigma(y)\cdot\lambda_{\pi,m,r,\sigma,n}}
{\pi(x)\cdot\bar\pi_{\pi,m,r}(y)}
\\
 &=&
\sum_{(x,\pi,m)\in\calX\times\Pi\times\calM,
\bar\pi_{\pi,m,r}(y)>0}
 p^{t-1}(\pi,m)\cdot
\frac{\gamma_{\pi,m,r}(x,y)\cdot\sigma(y)\cdot\lambda_{\pi,m,r,\sigma,n}}
{\bar\pi_{\pi,m,r}(y)}
\\
 &=&
\sigma(y)\cdot
\sum_{(\pi,m)\in\Pi\times\calM,
\bar\pi_{\pi,m,r}(y)>0}
\parend{
 p^{t-1}(\pi,m)\cdot\lambda_{\pi,m,r,\sigma,n}
\cdot
\sum_{x\in\calX}
\frac{\gamma_{\pi,m,r}(x,y)}
{\bar\pi_{\pi,m,r}(y)}
} 
\\
 &=&
\sigma(y)\cdot
\sum_{(\pi,m)\in\Pi\times\calM,
\bar\pi_{\pi,m,r}(y)>0}
 p^{t-1}(\pi,m)\cdot\lambda_{\pi,m,r,\sigma,n}
\\
 &=&
\sigma(y)\cdot
\sum_{(\pi,m)\in\Pi\times\calM}
 p^{t-1}(\pi,m)\cdot\lambda_{\pi,m,r,\sigma,n}
 \hskip 0.1in = \hskip 0.1in
\sigma(y)\cdot
 p^t(\sigma,n)
\end{eqnarray*}
}
which verifies claim 1 for $t$.
Claim 2 for $t$ follows trivially.

For the conclusion of the lemma, let $t > 0$, and let $r = r^t$.
We use claim 1 for $t-1$.
Recall that $\bar\pi_{\pi,m,r} = \sum_{\sigma\in\Pi, n\in\calM}
\lambda(\pi,m,r,\sigma,n)
\cdot \sigma $
for any full state $(\pi,m)$ of $\calA$.
Then
{\small
\begin{eqnarray*}
E\parend{\cost^t_{\tilde\calA}}
&=&
\sum_{\pi,\sigma\in\Pi,m,n\in\calM,x,y\in\calX}
q^{t-1}(x,\pi,m)\cdot
\Lambda_{x,\pi,m,r,y,\sigma,n}\cdot
\cost(x,r,y)\\
&=&
\sum_{\pi,\sigma\in\Pi,m,n\in\calM,x,y\in\calX\atop
\pi(x)>0,\sigma(y)>0}
p^{t-1}(\pi,m)\cdot\pi(x)
\cdot
\frac
{\gamma_{\pi,m,r}(x,y)
\cdot
\sigma(y)\cdot\lambda_{\pi,m,r,\sigma,n}
}
{\pi(x)
\cdot
\bar\pi_{\pi,m,r}(y)}
\cdot\cost(x,r,y)
\\
&=&
\sum_{\pi\in\Pi,m\in\calM,x,y\in\calX}
\parend{
p^{t-1}(\pi,m)\cdot
\gamma_{\pi,m,r}(x,y)\cdot\cost(x,r,y)
\cdot
\sum_{\sigma\in\Pi,n\in\calM,\sigma(y)>0}
\frac
{\lambda_{\pi,m,r,\sigma,n}\cdot\sigma(y)}
{\bar\pi_{\pi,m,r}(y)}
} 
\\
&=&
\sum_{\pi\in\Pi,m\in\calM,x,y\in\calX}
p^{t-1}(\pi,m)\cdot
\gamma_{\pi,m,r}(x,y)\cdot\cost(x,r,y)
\\
&=&
\sum_{\pi\in\Pi,m\in\calM}
\parend{
p^{t-1}(\pi,m)\cdot
\sum_{x,y\in\calX}\gamma_{\pi,m,r}(x,y)\cdot\cost(x,r,y)
}
\\
&=& \sum_{\pi\in\Pi,m\in\calM}
p^{t-1}(\pi,m)\cdot\cost_\calA(\pi,r,\bar\pi_{\pi,m,r})\\
&=& \sum_{\pi\in\Pi,m\in\calM}p^{t-1}(\pi,m)\cdot\cost_\calA(\pi,m,r)
\hskip 0.1in = \hskip 0.1in
E\parend{\cost^t_\calA}
\end{eqnarray*}
}
and we are done.
\end{proof}

\begin{theorem}\label{thm: mixed implies all standard}
If $\calA$ is a mixed model online algorithm for an online problem $\calP$,
there exist algorithms $\calA_1$, $\calA_2$, and $\calA_3$ for $\calP$,
of each of the standard models, such that, given any request sequence
$\varrho$, the cost (or expected cost) of each $\calA_i$ for $\varrho$ is no
greater than the cost (or expected cost) of $\calA$.
\end{theorem}

\begin{proof}
From Lemma \ref{lem: mixed yields beh}
and Theorem \ref{thm: all models equivalent}.
\end{proof}

\begin{corollary}\label{cor: mixed is comp imples comp}
If there is a $C$-competitive mixed model online algorithm for an
online problem $\calP$, there is a $C$-competitive online algorithm for
$\calP$ for each of the three standard models of
randomized online algorithms.
\end{corollary}

\section{Knowledge State Algorithms}
\label{sec: know}

We say that a function $\omega:\calX\to\reals$ is {\em Lipschitz\/} if
$\omega(y)\le  \omega(x)+d(x,y)$ for all $x,y\in\calX$.
An {\em estimator\/} is a non-negative Lipschitz function $\calX\to\reals$.
If $S\subseteq\calX$, we say that $S$
{\em supports an estimator \/} $\omega$ if, for any $y\in\calX$ there
exists some $x\in S$ such that $\omega(y)=  \omega(x)+d(x,y)$.
If $\omega$ is supported by a finite set, then there is a
unique minimal set $S$ which supports $\omega$, which
we call the {\em estimator support\/} of $\omega$.
(We use the term ``{\em support}" instead of ``estimator support"
if the context excludes ambiguity.) We note that all estimators
considered in this paper have finite support.
We say that an estimator $\omega$ has {\em zero minimum\/}
if $\min_{x\in\calX}\omega(x) = 0$. The next lemma
allows us to compare estimators
by examining finitely many values.

\begin{lemma}\label{lem: just check support}
Suppose $\omega$ and $\omega'$ are estimators,
and $S$ is the support of $\omega$.
Then $\omega(x)\ge\omega'(x)$ for all $x\in\calX$ if and only if
$\omega(y)\ge\omega'(y)$ for all $y\in S$.
\end{lemma}
\begin{proof}
One direction of the proof is trivial.
Suppose $\omega(x) < \omega'(x)$ and $\omega(y)\ge\omega'(y)$
for all $y\in S$.
Then there exists $y\in S$
such that $\omega(x)=\omega(y)+d(y,x)$.  It follows that
$\omega(y)=\omega(x)-d(y,x)<\omega'(x)-d(y,x)\le\omega'(y)$, contradiction.
\end{proof}

An example of an estimator is the {\em work function\/} of a request
sequence. If $x,y\in\calX$, we write $\cost_\malyopt^\varrho(x,y)$
to denote the minimal cost of servicing the request sequence $\varrho$
starting at configuration $x$ and ending at configuration $y$.
Then, if $\varrho$ is a request sequence, the {\em work function\/}
$\omega^\varrho:\calX\to\reals$ is defined by
$\omega^\varrho(x) = \cost_\malyopt(s^\malyz,\varrho,x)$.
If $\varrho$ is a request sequence, the {\em offset function\/} is
defined to be
$\bar\omega^\varrho = \omega^\varrho - \cost_\malyopt(\varrho)$,
a zero minimum estimator.
If $\omega$ is an estimator and if $r\in\calR$
is a request, we define function $\omega\update r$ as
$(\omega\update r)(y) = \min_{x\in\calX} \braced{\omega(x) + \cost(x,r,y)}$.
We call ``$\update$" the {\em update operator}. The following lemma
allows us to compute the update in finitely many steps.

\begin{lemma}\label{lem: update uses only support}
If $\omega$ is supported by $S$, then
$(\omega\update r)(y) = \min_{x\in S} \braced{\omega(x) + \cost(x,r,y)}$.
\end{lemma}
\begin{proof}
Trivially,
$(\omega\update r)(y) \le \min_{x\in S} \braced{\omega(x) + \cost(x,r,y)}$.
Pick $z\in\calX$ such that $(\omega\update r)(y)=\omega(z)+\cost(z,r,y)$.
Pick $x\in S$ such that $\omega(z) = \omega(x)+d(x,z)$.  Then
\begin{eqnarray*}
(\omega\update r)(y)
&=&
\omega(z)+\cost(z,r,y)
=
\omega(x)+d(x,z)+\cost(z,r,y)\\
&\ge&
\omega(x)+\cost(x,r,y)
\ge
(\omega\update r)(y)
\end{eqnarray*}
and we are done.
\end{proof}

We note that it is easy to verify that $\omega\update r$ is also an
estimator.  We briefly note the following lemma, which
is well-known (see, for example, \cite{ChrLar92B}).

\begin{lemma}\label{lem: update work}
If $\varrho = r^\malyone\ldots r^n$,
let $\varrho^t = r^\malyone\ldots r^t$ for all $t\le n$.  Then
 $\omega^\malyz(x)=d(s^\malyz,x)$ for all $x\in\calX$ and
$\omega^{\parend{\varrho^t}} = \omega^{\parend{\varrho^{t-1}}}\update r^t$
for all $t > 0$.
\end{lemma}
We use {\em estimators\/} and {\em adjustments\/} to analyze the
competitiveness of an online algorithm $\calA$.
More specifically, the combination of estimators and adjustments
allows us to estimate the optimal cost.
An online algorithm does not
know the optimal offline algorithm's cost at any given time,
but can keep track of the estimator, and use it as a guide.
The estimator is a real-valued function on configurations that is updated
at every step, and which estimates the cost of the optimal offline algorithm,
while the adjustment is a real number that is computed at every step.
Both the estimator and the adjustment may be calculated using randomization.

A {\em knowledge state algorithm\/} is a mixed online algorithm that
computes an adjustment and an estimator at each step, and uses the
current estimator as its memory state.  More formally, if
$\calA$ is a knowledge-state algorithm, then:
\begin{enumerate}
\item
At any given step, the full state of $\calA$ is a pair $(\pi,\omega)$,
where $\pi\in \Pi$ and $\omega:\calX\to\reals$
is the current estimator.  We call that pair the {\em current knowledge
state.}
\item
If $k=(\pi,\omega)$ is the knowledge state and the next request is $r$,
then $\calA$ computes an adjustment, a number which we call
$\offset_\calA(k,r)$, and uses randomization to pick a new knowledge state
$k'=(\pi',\omega')$.  More precisely, there are subsequent knowledge states
$k_i=(\pi_i,\omega_i)$ and subsequent weights $\lambda_i$ for $i=1,\ldots m$
such that
\begin{enumerate}
\item\label{eqn: las vegas update}
$(\omega\update r)(x) \ge \offset_\calA(k,r) +
   \sum^m_{i=1}\lambda_i\omega_i(x)$
for each $x\in\calX$.
\item
For each $i$, $\calA$ chooses $k'$ to be  $k_i$
with probability $\lambda_i$.
\item
  Let
 $\bar\pi=\sum^m_{i=1}\lambda_i\pi_i$.
 Define $\cost_\calA(k,r) = \cost(\pi,r,\bar\pi)$.
(As defined in the previous section in terms of the transportation problem)
\end{enumerate}

\item
Finally, if $\varrho=r^\malyone\ldots r^n$ is the
input request sequence, and the sequence of full states of $\calA$
is $k^\malyone\ldots k^n$, where $k^t=(\pi^t,\omega^t)$, we define
\begin{eqnarray*}
\cost^t_\calA(\varrho)&=&\cost_\calA\parend{k^{t-1},r^t}\mbox{ and }
\offset^t_\calA(\varrho)=\offset_\calA\parend{k^{t-1},r^t},\\
\cost_\calA(\varrho)&=&\sum^n_{t=1}\cost^t_\calA(\varrho) \mbox{ and }
\offset_\calA(\varrho)=\sum^n_{t=1}\offset^t_\calA(\varrho).
\end{eqnarray*}

\end{enumerate}

If $S\subseteq\calX$, we say that a knowledge state $(\pi,\omega)$
is {\em supported as a knowledge state} by $S$ if $\omega$ is
supported by $S$ (in the estimator sense) and $\pi$ is supported
(distributionally) by $S$.
Note that, in this case, $(\pi,\omega)$ can be represented by the finite
set of triples $\braced{(x,\pi(x),\omega(x))}_{x\in S}$.  We say that a
knowledge state algorithm {\em has finite support\/} if there is a
uniform bound on the cardinality of the supports of the knowledge states.
This bound is also called the {\em order\/} of the
knowledge state algorithm.

We say that {\em $\calA$ is $C$-competitive as a knowledge state algorithm\/}
if there is a constant $K$ such that
$E\parend{\cost_\calA(\varrho)}\le
C\cdot{}E\parend{\offset_\calA(\varrho)+\omega^n(x)}+K$
for any request sequence $\varrho = r^1\dots r^n$ and any $x\in\calX$.

\begin{lemma}\label{lem: is opt}
Given a request sequence
$\varrho = r^\malyone\ldots r^n$, then for all $x\in\calX$
\begin{eqnarray*}
E\parend{\omega^n(x)+\offset_\calA(\varrho)}
&\le&
\cost_\malyopt^\varrho(s^\malyz,x)
\label{eqn: total opt}
\end{eqnarray*}
\end{lemma}

\begin{proof}
Let
$s^\malyz=x^\malyz,x^\malyone,\ldots x^n=x\in\calX$ be the optimal service
of $\varrho$ that ends in $x$.  Thus:
 $\sum^n_{t=1}\cost(x^{t-1},r^t,x^t)
 =\cost_\malyopt(s^\malyz,x)$.
By (\ref{eqn: las vegas update}):
 $E\parend{\omega^t(x^t)+\offset^t_\calA(\varrho)}
 \le
 E\parend{\parend{\omega^{t-1}\update r^t}(x^t)}$ for all $t$.
 By definition: $E\parend{\parend{\omega^{t-1}\update r^t}(x^t)}
 \le
 E\parend{\omega^{t-1}(x^{t-1})}+\cost(x^{t-1},r^t,x^t)$
for all $t$.
Summing the inequalities over all $t$, and adding to the equation,
we obtain the result.
\end{proof}

\begin{lemma}
\label{lem: ks comp implies comp}
If a knowledge state algorithm $\calA$ is $C$-competitive
as a knowledge state algorithm, then $\calA$ is $C$-competitive.
\end{lemma}
\begin{proof}
Let $K$ be the constant given in the definition of $C$-competitiveness
for a knowledge state algorithm.
Let
$\varrho = r^\malyone\dots r^n$ be any request sequence, and let
 $s^\malyz=x^\malyz,x^\malyone,\ldots x^n\in\calX$ be the optimal service
of $\varrho$.
Since $\calA$ is $C$-competitive as a knowledge state algorithm:
\begin{eqnarray*}
E\parend{\cost_\calA(\varrho)}
&\le&
C\cdot E\parend{\offset_\calA(\varrho)}+
C\cdot E\parend{\omega^n(x^n))}+K\\
E\parend{\offset_\calA(\varrho)+\omega^n(x^n)}
&\le&
\cost_\malyopt(\varrho)~~~~~
{\mbox{\rm (by lemma \ref{lem: is opt})\ }}
\\
{\mbox{\rm We obtain:\ }}~~~~~~~~~~~~~~~~~~~~~~~
E\parend{\cost_\calA(\varrho)}
&\le&
C\cdot\cost_\malyopt(\varrho)+K
\end{eqnarray*}
\end{proof}

We now define a $C$-knowledge state potential  ({$C$-ks-potential\/},
for short) for a given knowledge state algorithm $\calA$.
Let $\Phi_{\calA}$ be a real-valued function on knowledge states.
Then we say that $\Phi_{\calA}$ is a {\em $C$-ks-potential for $\calA$\/}
if
\begin{enumerate}
\item
$\Phi_{\calA}(k)\ge 0$ for any $k$.
\item
If $k=(\pi,\omega)$ is the current knowledge state and $r$ is the next request,
$\braced{k_i=(\pi_i,\omega_i)}$ are the subsequents of that request, and
$\braced{\lambda_i}$ are the weights of the subsequents, let
$\Delta\Phi_{\calA}(k,r) =
\sum^m_{i=1}\lambda_i\Phi_{\calA}(\pi_i,\omega_i)-\Phi_{\calA}(\pi,\omega)$.
Then
\begin{eqnarray*}
 cost_\calA(k,r)+\Delta\Phi_{\calA}(k,r)&\le& C\cdot\offset_\calA(k,r).
\end{eqnarray*}
\end{enumerate}

\begin{theorem}\label{thm: ks pot implies comp}
If a knowledge state algorithm $\calA$ has a $C$-ks-potential,
then $\calA$ is $C$-competitive.
\end{theorem}

\begin{proof}
The proof follows easily from the definition
of a $C$-ks-potential and  Lemmas \ref{lem: is opt}
and \ref{lem: ks comp implies comp}
by straightforward arguments.
Let $\varrho = r^\malyone\ldots,\ldots r^n$ be a request sequence.
Let $k^\malyone, \ldots k^n$ be the sequence of knowledge states of $\calA$
given the input $\varrho$, where $k^t=(\pi^t,\omega^t)$.
Let $\Phi_{\calA}^t=\Phi_{\calA}(k^t)$, a random variable for each $t$.
Note that $\Phi_{\calA}^0$ is a constant.
Let $\Delta^t\Phi_{\calA}=\Delta\Phi_{\calA}(k^{t-1},r^t)$.
Note that $E(\Delta^t\Phi_{\calA}) = E(\Phi_{\calA}^t-\Phi_{\calA}^{t-1})$.
Let $x\in\calX$ be the configuration of the optimal algorithm
after $n$ steps.  Then
{\small
\begin{eqnarray*}
C\cdot\cost_\malyopt(\varrho)-E(\cost_\calA(\varrho))&\ge&\\
C\cdot
 E\parend{\omega^n(x)+\offset_\calA(\varrho)}-E(\cost_\calA(\varrho))&=&\\
C\cdot E\parend{\omega^n(x)+\sum^n_{t=1}\offset^t_\calA(\varrho)}
-E\parend{\sum^n_{t=1}\cost^t_\calA(\varrho)}&=&\\
E\parend{C\cdot\omega^n(x)+\sum^n_{t=1}\parend{C\cdot\offset^t_\calA(\varrho)
-\cost^t_\calA(\varrho)}}&=&\\
E\parend{C\cdot\omega^n(x)
+\Phi_{\calA}^n+\sum^n_{t=1}\parend{C\cdot\offset^t_\calA(\varrho)
-\cost^t_\calA(\varrho)
-\Delta^t\Phi_{\calA}}}-\Phi_{\calA}^0&\ge&\\
E\parend{C\cdot\omega^n(x)+\Phi_{\calA}^n}-\Phi_{\calA}^0&\ge&-\Phi_{\calA}^0
\end{eqnarray*}
}
The first inequality above is from Lemma \ref{lem: is opt}.
The last two inequalities are from the definition of a  $C$-ks-potential.
It follows that
$E\parend{\cost_\calA(\varrho)}\le
C\cdot\cost_\malyopt(\varrho)+\Phi_{\calA}^0$,
and, by Lemma \ref{lem: ks comp implies comp},
 we are done.
\end{proof}

We can define a {\em forgiveness\/} online algorithm to be a knowledge
state algorithm with the special restriction that there is always
exactly one subsequent. We note that historically, forgiveness came first,
so we can think of the knowledge state approach as being a generalization
of forgiveness.  A forgiveness algorithm can be deterministic,
such as EQUIPOISE, a deterministic online 11-competitive algorithm for the
3-server problem (that was the best known competitiveness for that problem
at that time), or distributional, such as EQUITABLE, an $H_k$-competitive
distributional online algorithm for the $k$-cache problem.
(See \cite{AcChNo00,ChrLar92C}.)


\section{Knowledge State Algorithms for the
Cache Problem}\label{sec: ks cache}
We now consider the $k$-cache problem for fixed $k\ge 2$.
The $k$-cache problem reduces to online optimization, as defined in Section
\ref{sec: online opt prob} of this paper, as follows:
\begin{enumerate}
\item
There is a set of {\em pages\/}.
\item
$\calX$ is the set of all $k$-tuples of distinct pages.
If the configuration of an algorithm is $x\in\calX$,
that means that the pages
that constitute $x$ are in the cache.
\item
The initial configuration is the initial cache.
\item
If $x,y\in\calX$, then $d(x,y)$ is the cost of
changing the cache from $x$ to $y$.
Since we assume that it costs 1 to eject a page and bring in a new page,
$d(x,y)$ is the cardinality of the set $x-y$.
\item
$\calR$ is simply the set of all pages.
If a page $r$ is requested, it means that the algorithm must ensure that
$r$ is in the cache at some point as it moves between configurations.
Thus, for any $x,y\in\calX$ and any $r\in\calR$, we have
{\small
\begin{eqnarray*}
\cost(x,r,y)=\leftbracedthree
{2\mbox{\rm\ if\ }x=y,\ r\not\in x}
{d(x,y){\rm\ if\ }r\in x\mbox{\rm\ or\ }r\in y}
{d(x,y)+1\mbox{\rm\ otherwise}}
\end{eqnarray*}
}
\end{enumerate}
To complete the reduction, we observe that the support of any configuration
request pair $(x,r)$ is finite.  If $r\in x$, that support has only one
element, namely $x$, while otherwise, it has $k$ elements, namely
$\braced{x-a+r\ |\ a\in x}$.

\myparagraph{Bar Notation for the Cache Problem.}
We introduce a convenient notation, a modification of the bar notation of
Koutsoupias and Papadimitriou \cite{KouPap94B},
for offset functions for the $k$-cache problem, which we call the
{\em bar notation\/}.\footnote{The notation of \cite{KouPap94B} differs
slightly from that given here, although it is based on the same concept.}
Let $\alpha$ be a string consisting of at least $k$ page names and exactly
$k$ bars, with the condition that at least $i$ page names are to the
left of the $i^\tH$ bar.  Then $\alpha$ defines an offset function $\omega$
as follows.  Let $S\subseteq\calX$ be the set of all configurations $x$
such that, for each
$i = 1,\ldots k$, the names of at least $i$ members of $x$ are written
to the left of the $i^\tH$ bar.  Let $\omega$ be the estimator such
that $S$ is the support of $\omega$, and such that $\omega(x) = 0$
for each $x\in S$.
For example for $k=2$, $ab||$ denotes the estimator whose support
consists of just the configuration $\braced{a,b}$,
and which takes the value zero on
that configuration.
  For $k=4$,
$ab||cd|ef|$ denotes the estimator whose support consists of the
configurations $\braced{a,b,c,d}$, $\braced{a,b,c,e}$, $\braced{a,b,c,f}$,
$\braced{a,b,d,e}$, and $\braced{a,b,d,f}$, and which takes the value zero on
those configurations.
From \cite{KouPap94B}, we have:
\begin{lemma}\label{lem: bar lemma}
A function $\omega$ is an offset function for the $k$-cache problem
if and only if it can be expressed using the bar notation.
\end{lemma}

\subsection{A $\frac{3}{2}$-Competitive Knowledge State Algorithm
for the 2-Cache Problem}\label{subsec: k2}

Recall that PARTITION (introduced in \cite{KouPap94B})
 is optimally competitive for the $k$-cache problem,
but uses unbounded memory to achieve the optimal competitiveness of $H_k$.
The memory state of PARTITION is, in fact, the classic offset function,
which, in the worst case, requires keeping track of every past request.
We now show how the use of knowledge states simplifies the definition,
and in fact the memory requirement, of an optimally competitive
randomized algorithm for the 2-cache problem, which we call $K_2$.

\begin{figure}[ht!]
\centerline{\epsfig{file=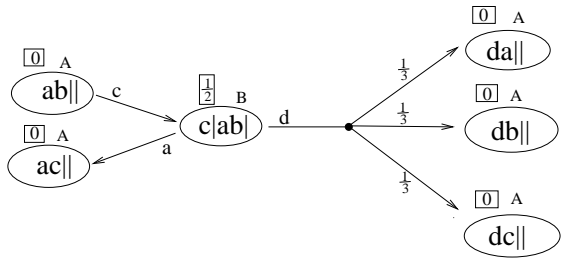, width=2.9in}}
\caption{Schematic for the $2$-Cache Knowledge State Algorithm}
\label{fig: k2}
\end{figure}

\myparagraph{Knowledge States of $K_2$.}
We will follow the rule that, at each step, the adjustment is
as large as possible, so that the minimum of the estimator will always
be zero.  This guarantees that any potential will always be non-negative.
If there are infinitely many pages,
$K_2$ has infinitely many knowledge states, but, up to
symmetry, it has only two.
Each such knowledge state of $K_2$ is supported by a set of cardinality at
most 2, hence has at most three active pages, and therefore its equivalent
behavioral algorithm has at most one bookmark.

In the definitions given below, we say that
two pages to are {\em equivalent\/} for a given knowledge state
if they can be transposed without changing the knowledge state.
\vskip 0.2in

\begin{enumerate}
\item
If $a,b$ are pages, let
 $A^{a,b}=\parend{\braced{a,b},ab||}$.
In this case, $a$ and $b$ are equivalent, \ie, $A^{a,b} = A^{b,a}$.
\item
If $a,b,c$ are pages, let
$B^{a,b,c}=\parend{\onehalf\braced{a,b}+\onehalf\braced{a,c},a|bc|}$,
where $\onehalf\braced{a,b}+\onehalf\braced{a,c}$ denotes
the distribution which is $\onehalf$ on the configuration $\braced{a,b}$
and $\onehalf$ on the configuration $\braced{a,c}$.
In this case $b$ and $c$ are equivalent, \ie, $B^{a,b,c} = B^{a,c,b}$.
\end{enumerate}

We list below the action of $K_2$.
In each case, $a,b,c,d$ are distinct pages.

\begin{enumerate}[1.]
\item
If $\braced{a,b}$ is the initial cache,
the initial knowledge state is $A^{a,b}$.

\item
If the current knowledge state is $A^{a,b}$ then
\begin{enumerate}[(i)] 
\item \label{act: k2 trivial A}
if the request is $a$, 
the new knowledge state is $A^{a,b}$. 

\item\label{act: k2 Ac}
if the request is
$c$, then the new knowledge state is $B^{c,a,b}$.
\end{enumerate}

\item
If the current knowledge state is $B^{a,b,c}$ then
\begin{enumerate}[(i)]
\item 
\label{act: k2 trivial B}
if the new request
is $a$, the new knowledge state is $B^{a,b,c}$.
\item\label{act: k2 Bb}
if the new request
is $b$, the new knowledge state is $A^{b,a}$.
\item\label{act: k2 Bd}
if the new request is
$d\notin\braced{a,b,c}$, then there are three subsequents, namely
$A^{d,a}$, $A^{d,b}$, $A^{d,c}$.  The distribution
on the subsequents is uniform, \ie, each is chosen with probability
$\onethird$.
\end{enumerate}
\end{enumerate}

Actions \ref{act: k2 trivial A} and \ref{act: k2 trivial B}
are requests to the first block of pages, in the sense of the bar notation.
Since the bar notation implies that each page in
the first block can be assumed to be in the cache, such a request
is ignored by any sensible online algorithm, which means, in our case,
that the estimator is unchanged and the adjustment is zero.
We call such requests {\em trivial\/}.

We define a potential
$\Phi$ by  $\Phi(A^{a,b})=0$ and $\Phi(B^{a,b,c})=\onehalf$.

\begin{lemma}\label{lem: pot for k2}
$\Phi$ is a $\threehalves$-ks-potential for $K_2$.
\end{lemma}
\begin{proof}
Let $k$ be the current knowledge state and $r$ the new request.
Write $\Delta\Phi$ for increase in potential in the given step.
We will show that
\begin{eqnarray}
 \cost+\Delta\Phi &\le& \threehalves\offset
\label{eqn: k2 pot}
\end{eqnarray}
in all cases.
In trivial actions, namely Cases
 \ref{act: k2 trivial A} and \ref{act: k2 trivial B},
$\cost=\Delta\Phi=\offset$, and we are done.

We first note that:
\begin{eqnarray*}
ab||\update c &=& c|ab|+1\\
a|bc|\update b &=& ab||\\
a|bc|\update d &\ge&
\onethird da||+\onethird db||+\onethird dc||+\onethird
\end{eqnarray*}
By Lemma \ref{lem: just check support},
the last inequality need only be verified for configurations
in $\braced{\braced{d,a},\braced{d,b},\braced{d,c}}$, the support
set of $a|bc||\update d$.

\mycase{Action \ref{act: k2 Ac}}
 In this case $k=A^{a,b}$ and $r$ is a new page, $c$.

$ab||\update c = c|ab|+1$.
thus $\offset = 1$.
Since the algorithm must bring in a new page,
and since the probability is zero that the minimum transport brings
in any other page, $\cost = 1$.
$\Delta\Phi = \onehalf$, and we are done.

\mycase{Action \ref{act: k2 Bb}},
\ie, $k=B^{a,b,c}$ and $r=b$.

Recall $a|bc|\update b = ab||$.
Note that $\offset = 0$, since, as functions, $ab|| \ge a|bc|$
on the set of all configurations.
$\cost=\onehalf$, since the probability is $\onehalf$ that the algorithm
does nothing, and the probability is $\onehalf$ that it ejects $c$ and
brings in $b$.  $\Delta\Phi= -\onehalf$, and we are done.

\mycase{Action \ref{act: k2 Bd}},
\ie, $k=B^{a,b,c}$ and $r$ is a new page, $d$.

Recall $a|bc|\update d\ge\onethird da||+
\onethird db||+\onethird dc||+\onethird$, thus
$\offset=\onethird$.
Since the algorithm must bring in a new page,
and since the probability is zero that the minimum transport brings
in any other page, $\cost = 1$.
$\Delta\Phi = -\onehalf$, and we are done.

This completes the proof of all cases.
\end{proof}

We have:

\begin{corollary}\label{thm: k2 comp pot}
$K_2$ is $\threehalves$-competitive.
\end{corollary}

We note that the number of active pages, \ie, pages contained in a
support configuration, is never more
than three.  The number three is minimal, as given by the theorem below:
\begin{theorem}\label{thm: at least 3 active pages}
There is no knowledge state algorithm for the 2-cache
problem that is $\frac{3}{2}$-competitive as a knowledge state algorithm,
and which never has more than two active pages, \ie, no bookmarks.
\end{theorem}
\begin{proof}
If a knowledge state algorithm for the 2-cache problem never has more than
two active pages, then it can have no bookmarks, hence is trackless.
By Theorem 2 of \cite{BeFlLa00}, there is no $\threehalves$-competitive
trackless online algorithm for the 2-cache problem.
\end{proof}

\subsection{An Optimally Competitive Knowledge State Algorithm
for the 3-Cache Problem}

We define a knowledge state algorithm $K_3$ which is $H_3$-competitive
for the 3-cache problem.  Recall that $H_3=\frac{11}{6}$.
Up to symmetry, $K_3$ has six knowledge states.  The number of active
pages, \ie, pages contained in a support configuration, is never more
than five.

The knowledge states of $K_3$ will be defined as follows.
As in the case of $K_2$,
We say that two pages are {\em equivalent\/} if they can be
transposed without changing the knowledge state.
\begin{enumerate}
\item
$A^{a,b,c}=\parend{\braced{a,b,c},abc|||}$ for any three pages $a,b,c$.
The pages $a$, $b$, and $c$ are all equivalent,
\ie, $A^{a,b,c} = A^{b,a,c} = A^{a,c,b}$, \etc
\item
$B^{a,b,c,d}=\parend{\onethird\braced{a,b,c}
+\onethird\braced{a,b,d}+\onethird\braced{a,c,d},
a|bcd||}$ for any four pages $a,b,c,d$.
The pages $b$, $c$, and $d$ are all equivalent.
\item
$C^{a,b,c,d}=\parend{\onehalf\braced{a,b,c}
+\onehalf\braced{a,b,d},ab||cd}$ for any four pages $a,b,c,d$.
The pages $a$ and $b$ are equivalent, and $c$ and $d$ are equivalent.
\item
$D^{a,b,c,d,e}=
\parend{
\onesixth\braced{a,b,c}+\onesixth\braced{a,b,d}
+\onesixth\braced{a,b,e}+\onesixth\braced{a,c,d}
+\onesixth\braced{a,c,e}+\onesixth\braced{ade},a|bcde||}$
for any five pages $a,b,c,d,e$.  The pages $b,c,d,e$ are equivalent.
\item
$E^{a,b,c,d,e}=\parend{\onehalf\braced{a,b,c}
+\onefourth\braced{a,b,d}+\onefourth\braced{a,b,e},ab||cde|}$
for any five pages $a,b,c,d,e$.
The pages $a$ and $b$ are equivalent, and $d$ and $e$ are equivalent.
\item
$F^{a,b,c,d,e}=\parend{\onehalf\braced{a,b,c}
+\oneeighth\braced{a,b,d}+\oneeighth\braced{a,b,e}
+\oneeighth\braced{a,c,d} +\oneeighth\braced{a,c,e},a|bc|de|}$
for any five pages $a,b,c,d,e$.
The pages $b$ and $c$ are equivalent, and $d$ and $e$ are equivalent.

 \newcounter{jump}
 \setcounter{jump}{\value{enumi}}
\end{enumerate}

\begin{figure}[ht!]
\centerline{\epsfig{file=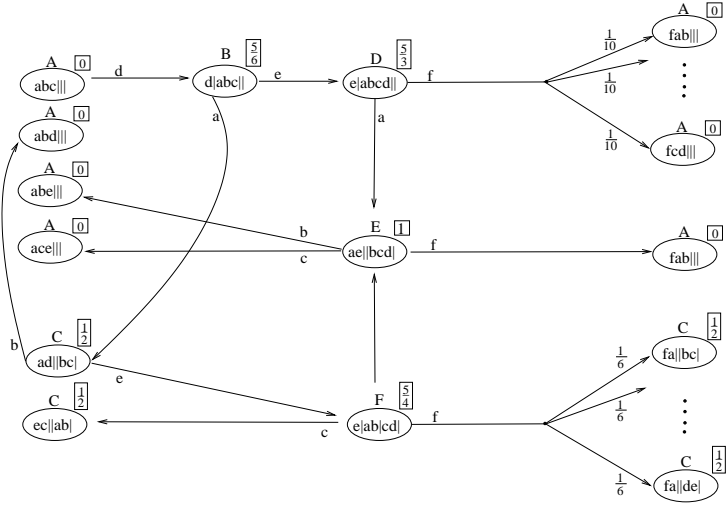, width=3.8in}}
\caption{Schematic for the $3$-Cache Knowledge State Algorithm}
\label{fig: k3}
\end{figure}

The actions are of $K_3$ are formally defined below.
In each case, $a,b,c,d,e,f$ are distinct pages.  We do not need
to consider separate cases for requests to pages which are equivalent.

\begin{enumerate}[1.]
\item\label{act: k3 initial A}
If $\braced{a,b,c}$ is the initial cache,
the initial knowledge state is $A^{a,b,c}$.

\item
If the current knowledge state is $A^{a,b,c}$ then
\begin{enumerate}[(i)]
\item \label{act: k3 Aa}
if the new request is $a$, the new knowledge state is $A^{a,b,c}$.

\item\label{act: k3 Ad}
if the new request is
some page $d\notin\braced{a,b,c}$, the new knowledge state is
$B^{d,a,b,c}$.
\end{enumerate}

\item
If the current knowledge state is $B^{a,b,c,d}$ then
\begin{enumerate}[(i)]
\item
\label{act: k3 Ba}
if the new request
is $a$, the new knowledge state is $B^{a,b,c,d}$.

\item\label{act: k3 Bb}
if the new request
is $b$, the new knowledge state is $C^{a,b,c,d}$.

\item\label{act: k3 Be}
if new request is some
page $e\notin\braced{a,b,c,d}$, the new knowledge state is $D^{e,a,b,c,d}$.
\end{enumerate}

\item
If the current knowledge state is $C^{a,b,c,d}$ then

\begin{enumerate}[(i)]
\item \label{act: k3 Ca}
if the new request is $a$, the new knowledge state is $C^{a,b,c,d}$.

\item\label{act: k3 Cc}
if the new request
is $c$, the new knowledge state is $A^{a,b,c}$.

\item\label{act: k3 Ce}
if the new request is some
page $e\notin\braced{a,b,c,d}$, the new knowledge state is $F^{e,a,b,c,d}$.
\end{enumerate}

\item
If the current knowledge state is $D^{a,b,c,d,e}$ then 
\begin{enumerate}[(i)]
\item \label{act: k3 Da} 
if the new request is $a$,
the new knowledge state is $D^{a,b,c,d,e}$.

\item\label{act: k3 Db}
if the new request is $b$,
the new knowledge state is $E^{a,b,c,d,e}$%

\item\label{act: k3 Df}
if the new request is
some page $f\notin\braced{a,b,c,d,e}$, then the new knowledge state is
chosen uniformly from among the following ten knowledge states:
$A^{abc}$, $A^{abd}$, $A^{abe}$, $A^{abf}$, $A^{acd}$, $A^{ace}$,
$A^{acf}$, $A^{ade}$, $A^{adf}$, and $A^{aef}$.
\end{enumerate}

\item
If the current knowledge state is $E^{a,b,c,d,e}$ then
\begin{enumerate}[(i)]
\item
\label{act: k3 Ea}
if the new request is $a$,
the new knowledge state is $E^{a,b,c,d,e}$.

\item\label{act: k3 Ec}
if the new request is $c$,
the new knowledge state is $A^{a,b,c}$.

\item\label{act: k3 Ed}
if the new request is $d$,
the new knowledge state is $A^{a,b,d}$.

\item\label{act: k3 Ef}
if the new request is
some page $f\notin\braced{a,b,c,d,e}$, then the new knowledge state is
$A^{f,a,b}$.
\end{enumerate}

\item
If the current knowledge state is $F^{a,b,c,d,e}$ then
\begin{enumerate}[(i)]
\item\label{act: k3 Fa}
if the new request is $a$,
the new knowledge state is $F^{a,b,c,d,e}$.

\item\label{act: k3 Fb}
if the new request is $b$,
the new knowledge state is $E^{a,b,c,d,e}$.

\item\label{act: k3 Fd}
if the new request is $d$,
the new knowledge state is $C^{a,d,b,c}$.

\item\label{act: k3 Ff}
if the new request is
some page $f\notin\braced{a,b,c,d,e}$, the new knowledge state is
chosen uniformly from among the following six knowledge states:
$C^{f,a,b,c}$, $C^{f,b,a,c}$, $C^{f,c,a,b}$,
$C^{f,a,d,e}$, $C^{f,b,d,e}$, and $C^{f,c,d,e}$.
\end{enumerate}
\end{enumerate}

We define a potential $\Phi$ on the knowledge states as follows:
$\Phi(A^{a,b,c})=0$,
$\Phi(B^{a,b,c,d})=\fivesixths$,
$\Phi(C^{a,b,c,d})=\onehalf$,
$\Phi(D^{a,b,c,d,e})=\onehalf$,
$\Phi(E^{a,b,c,d,e})=1$, and
$\Phi(F^{a,b,c,d,e})=\fivefourths$.

\begin{lemma}\label{lem: pot for k3}
$\Phi$ is an $\elevensixths$-ks-potential for $K_3$.
\end{lemma}
\begin{proof}
For each action of $K_3$, let $\Delta\Phi$ be the increase in potential.
We will show that
\begin{eqnarray}
\cost+\Delta\Phi &\le&\elevensixths\offset
\label{eqn: k3 pot}
\end{eqnarray}
In each case, the value of $\Delta\Phi$ can be computed by
simple subtraction.  We need only compute the values of $\cost$
and $\offset$ for each action, after which the inequality (\ref{eqn: k3 pot})
follows by simple arithmetic.

\mycase{Actions \ref{act: k3 Aa}, \ref{act: k3 Ba}, \ref{act: k3 Ca},
\ref{act: k3 Da}, \ref{act: k3 Ea}, \ref{act: k3 Fa}}.
These actions are trivial, and thus
$\offset = \cost = \Delta\Phi = 0$, and we are done.

\mycase{Actions \ref{act: k3 Ad}, \ref{act: k3 Be}, \ref{act: k3 Ce}}.
In these actions, the request is to a new page, and the probability that
any other page is in the cache after the action does not increase: thus
$\cost = 1$.
We also know that $\offset = 1$ because
\begin{eqnarray*}
abc|||\update d &=& d|abc||+1\\
a|bcd||\update e &=& e|abcd||+1\\
ab||cd|\update e &=& e|ab|cd|+1
\end{eqnarray*}
The remainder of the verification of (\ref{eqn: k3 pot}) for each
of those actions consists of simple arithmetic.

\mycase{Actions \ref{act: k3 Ec} and \ref{act: k3 Ed}}.
Note that $\offset = 0$ since
\begin{eqnarray*}
ab|cde||\update c &=& abc|||\\
ab|cde||\update d &=& abd|||
\end{eqnarray*}
In each case, we must keep $a$ and $b$ and
eject the other two unrequested pages.  The probability is $\onehalf$
that $c$ is in our cache, and $\onefourth$ that $d$ is in our cache,
thus $\cost = \onehalf$ for Action \ref{act: k3 Ec},
and $\cost = \threefourths$ for Action \ref{act: k3 Ed}.
Since $\Delta\Phi = -\onehalf$ for both actions, we are done.

\mycase{\ref{act: k3 Db} and \ref{act: k3 Fb}}.
Note that $\offset = 0$ since
\begin{eqnarray*}
a|bcde||\update b &=& ab|cde||\\
a|bc|de|\update b &=& ab|cde||
\end{eqnarray*}
For Action \ref{act: k3 Db}, recall that the distribution of $D^{a,b,c,d,e}$
is uniform on six configurations.
To compute $\cost$, we describe
a minimal transport between the distribution of $D^{a,b,c,d,e}$ and
the distribution of $E^{a,b,c,d,e}$. That transport  is defined as follows:
\begin{description}
\item If the previous configuration is $\braced{a,b,c}$, $\braced{a,b,d}$,
or $\braced{a,b,e}$, do nothing.
\item If the previous configuration is $\braced{a,c,d}$, eject $d$.
\item If the previous configuration is $\braced{a,c,e}$, eject $e$.
\item If the previous configuration is $\braced{a,d,e}$, eject $d$ with
probability $\onehalf$, and eject $e$ with probability $\onehalf$.
\end{description}
Thus, $\cost = \onehalf$.  It is a routine verification that the required
distribution for $E^{a,b,c,d,e}$ is achieved.  Since
$\Delta\Phi = -\twothirds$, we have verified (\ref{eqn: k3 pot}) for
Action \ref{act: k3 Db}.

For Action \ref{act: k3 Fb}, recall that the distribution of $F^{a,b,c,d,e}$
is $\onehalf$ on $\braced{a,b,c}$, and is $\oneeighth$ on each of
$\braced{a,b,d}$, $\braced{a,b,e}$, $\braced{a,c,d}$, and $\braced{a,c,e}$.
A minimal transport can be defined as follows:
if $b$ is already in the cache we do nothing, while otherwise, we eject $c$.
Thus, $\cost = \onefourth$.  It is a routine verification that the required
distribution for $E^{a,b,c,d,e}$ is achieved.
Since $\Delta\Phi = -\onefourth$, we have verified (\ref{eqn: k3 pot}) for
Action \ref{act: k3 Fb}.

\mycase{Actions \ref{act: k3 Bb}, \ref{act: k3 Cc}, \ref{act: k3 Fd}}.
Note that $\offset = 0$ since
\begin{eqnarray*}
a|bcd||\update b &=& ab|cd||\\
ab||cd|\update c &=& abc|||\\
a|bc|de|\update d &=& ad||bc|\\
\end{eqnarray*}

For Action \ref{act: k3 Bb}, recall that the distribution of $B^{a,b,c,d}$
is uniform on $\braced{a,b,c}$, $\braced{a,b,d}$, and $\braced{a,c,d}$.
If $b$ is already in the cache we do nothing, while otherwise, we eject $c$
with probability $\onehalf$ and eject $d$ with probability $\onehalf$.
Thus, $\cost = \onethird$.  It is a routine verification that the required
distribution for $C^{a,b,c,d}$ is achieved.
Since $\Delta\Phi = -\onethird$, we have verified (\ref{eqn: k3 pot}) for
Action \ref{act: k3 Bb}.

For Action \ref{act: k3 Cc}, recall that the distribution of $C^{a,b,c,d}$
is uniform on $\braced{a,b,c}$ and $\braced{a,b,d}$.
If $c$ is already in the cache we do nothing, while otherwise, we eject $d$.
Thus, $\cost = \onehalf$.  The resulting distribution is
concentrated at $\braced{a,b,c}$, as required for the knowledge state
$A^{a,b,c}$.
Since $\Delta\Phi = -\onehalf$, we have verified (\ref{eqn: k3 pot}) for
Action \ref{act: k3 Cc}.

For Action \ref{act: k3 Fd}, recall that the distribution of $F^{a,b,c,d,e}$
is $\onehalf$ on $\braced{a,b,c}$, and is $\oneeighth$ on each of
$\braced{a,b,d}$, $\braced{a,b,e}$, $\braced{a,c,d}$, and $\braced{a,c,e}$.
If $d$ is already in the cache we do nothing.  If $e$ is in the cache,
we eject $e$.  Otherwise, the cache must be $\braced{a,b,c}$, in which
case we eject $b$ or $c$ with equal probability.
Thus, $\cost = \threefourths$.  It is a routine verification that the required
distribution for $C^{a,d,b,c}$ is achieved.
Since $\Delta\Phi = -\threefourths$, we have verified (\ref{eqn: k3 pot}) for
Action \ref{act: k3 Fd}.

\mycase{Action \ref{act: k3 Ef}}.
Note that $\offset \ge 0$ since $ab||cde|\update f = f|ab|cde|+1\ge abf|||$.
By Lemma \ref{lem: just check support},
this inequality need only be verified for the configurations
in the support of $ab||cde|\update f$.
Whatever the initial configuration is, $a$ and $b$ are in the cache.
Simply eject the other page.  Thus, $\cost=1$.  $\Delta\Phi = -1$,
and we are done.

\mycase{Actions \ref{act: k3 Df}, \ref{act: k3 Ff}}.
Let 
\[\omega^{Df} = \onetenth\parend{fab|||+fac|||+fad|||+fae|||
+fbc|||+fbd|||+fbe|||+fcd|||+fde|||},
\] 
and
let 
\[\omega^{Ff} = \onesixth\parend{fa||bc|+fb||ac|+fc||ab|
+fa||de|+fb||de|+fc||de|}.
\]
We note:
\begin{eqnarray*}
a|bcde||\update f = f|abcde|| + 1 &\ge& \omega^{Df}\\
a|bc|de|\update f = f|abc|de|+1 &\ge& \omega^{Ff} - \onesixth
\end{eqnarray*}
By Lemma \ref{lem: just check support},
these inequalities need only be verified for the configurations
in the support of $a|bcde||\update f$ and $a|bc|de|\update f$,
respectively.
We thus have $\offset \ge 0$ for \ref{act: k3 Df},
and $\offset \ge \onesixth$ for \ref{act: k3 Ff}.

To compute $\cost$, we give minimal transportations from the distribution of
$D^{a,b,c,d,e}$, respectively $F^{a,b,c,d,e}$, to
the weighted sum of distributions of the subsequents, for each of the two
cases.
For Action \ref{act: k3 Df},
whatever the initial configuration is, $a$ is in the cache.
Eject $a$ with probability $\threefifths$, and eject each of the
other two pages with probability $\onefifth$ each.
  It is a routine verification that the required
distribution is achieved.  Thus, $\cost=1$.
  $\Delta\Phi = -\fivethirds$, and we are done.

For Action \ref{act: k3 Ff}, the probability is $\onehalf$ that the
initial configuration is $\braced{a,b,c}$.  In this case, eject one
of the three pages, each with probability $\onethird$.
Otherwise, the cache will contain $a$, and either $b$ or $c$ but not
both:  eject $a$ with probability $\twothirds$, and otherwise eject
either $b$ or $c$.   It is a routine verification that the required
distribution is achieved.  Thus, $\cost=1$.
  $\Delta\Phi = -\threefourths$, and we are done.

This completes the proof of all cases.
\end{proof}

\begin{corollary}\label{thm: k3 comp pot} 
$K_3$ is $\elevensixths$-competitive. 
\end{corollary} 

\section{Experimental Work and the Server Problem}\label{sec: server}

It is our hope that our technique will yield an order 2 knowledge state
algorithm whose competitiveness is provably less than 2 for all metric
spaces.  

We  mention briefly progress
by giving results for a class of is ``one step up" in complexity
from the class of uniform metric spaces. 
We consider the class of metric spaces $M_{2,4}$,
which consists of all metric spaces where every distance is
either 1 or 2, and where the perimeter of every triangle is either 3 or 4.
(The classic octahedral graph, which has six points,
is a member of this class, as defined by Schl{\"a}fli \cite{Schl57}.)
We have a computer generated order 2 knowledge state algorithm for the
2-server problem in this class: its competitiveness is $\frac{7}{4}$.
We note that we also have calculated 
(through computer experimentation) the minimum value of $C$
in the sense that no lower competitiveness for any order 2
knowledge state algorithm for $M_{2,4}$
can be proved using the methods described here.
This value is $C=\frac{173+\sqrt{137}}{112}$.
We briefly mention that there is an order 3 knowledge state algorithm
for $M_{2,4}$ which has, up to equivalence, only seven knowledge states,
and is $\frac{19}{12}$-competitive.  We also can prove that no randomized
online algorithm for the 2-server problem for $M_{2,4}$ can achieve
competitiveness less than $\frac{19}{12}$.  All knowledge states and
probabilities in this order 3 algorithm can be described using
only rational numbers.  

These results, as well as our
results for the server problem in uniform spaces
(equivalent to the caching problem), indicate a natural trade-off between
competitiveness and memory of online randomized algorithms.

\end{document}